# System Model-Based Definition of Modeling Language Semantics

Hans Grönniger, Jan Oliver Ringert, and Bernhard Rumpe

Lehrstuhl Informatik 3 (Softwaretechnik), RWTH Aachen, Germany

**Abstract.** In this paper, we present an approach to define the semantics for object-oriented modeling languages. One important property of this semantics is to support underspecified and incomplete models. To this end, semantics is given as predicates over elements of the semantic domain. This domain is called the system model which is a general declarative characterization of object systems. The system model is very detailed since it captures various relevant structural, behavioral, and interaction aspects. This allows us to re-use the system model as a domain for various kinds of object-oriented modeling languages. As a major consequence, the integration of language semantics is straight-forward. The whole approach is supported by tools that do not constrain the semantics definition's expressiveness and flexibility while making it machine-checkable.

## 1 Introduction

Modeling is an integral part of complex software system development projects. The purpose of models ranges from assisting developers and customers communicate to test case generation or (automatic) derivation of the developed system. A prominent example modeling language is UML [1]. Actually, it is a family of languages used to model various aspects of a software system. While UML is widely used, domain specific modeling languages emerged recently that allow developers and even customers to express solutions to well-defined problems in a concise way.

A complete definition of a modeling language consists of the description of its syntax, including well-formedness rules and its semantics (meaning) [2]. It is widely accepted that a commonly agreed formal semantics of a language is advantageous because it avoids problems like misunderstandings between people and lack of interoperability between tools. Additionally, semantics can also be used to formally reason about system properties for verification purposes. However, many languages are often specified through their syntax only and lack a precise semantics beyond informal explanations. UML is again a prominent example which has been standardized without a formal semantics, even though debate has started more than ten years ago [3,4].

Various efforts for the definition of a formal semantics for a modeling language like UML have shown that this really is a difficult task for the following reasons:

- Multiple views and multiple models describe overlapping parts of the system. Thus, fundamentally different modelling concepts for structure, behavior and interaction have to be given an integrated semantics.
- As opposed to programming language semantics, modeling languages are used for specification. In particular high-level, abstract models are not necessarily executable. Instead, models tend to be incomplete and underspecified and thus their semantics must allow underspecification. A semantic definition has to provide a meaning for those models that cannot be described as an execution.
- The semantics has to be precise but not completely fixed. In UML terms, it should support semantic variation points that allow different stakeholders to provide a specialized interpretation for certain constructs.

Although UML is currently one of our main targets, the approach presented in this paper is not restricted to UML. Instead, the process of defining the semantics of a modeling language might even be more important for newly defined domain specific languages since it guides developers through the task of developing a formal semantics.

This paper presents our approach to define the semantics of object-oriented modeling languages which explicitly addresses the challenges mentioned above. The rest of the paper is structured as follows. Sect. 2 discusses our approach in general and motivates the usage of a single semantic domain that was carefully defined to capture the most important concepts of object-oriented systems. This domain is introduced in greater detail in Sect. 3 which also presents an implementation in the theorem prover Isabelle/HOL [5] as part of the proposed tool support. Sect. 4 is concerned with the precise definition of the syntax of a language using the framework MontiCore [6]. Furthermore, an automatic derivation of the abstract syntax as an Isabelle/HOL data type is outlined. With syntax and semantic domain specified and implemented in Isabelle/HOL, the process of defining the semantic mapping is described in Sect. 5. The mapping again is formalized in Isabelle/HOL. A running example is used throughout the paper for which a short verification application is also presented in Sect. 5. Related work is discussed in Sect. 6 and conclusions are drawn in Sect. 7.

## 2 General Approach

As indicated in Fig. 1, the semantics of a modeling language consists of the following basic parts [7]:

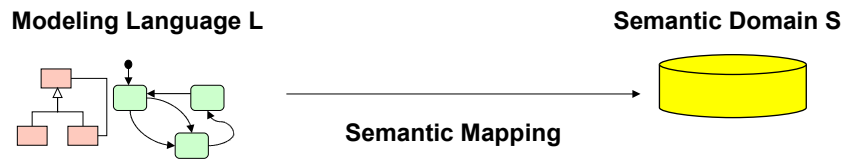

**Fig. 1.** Basic parts of a semantics definition

- the syntax of the language in question L – be it graphical or textual,
- the semantic domain S, a domain well-known and understood based on a well-defined mathematical theory, and
- the semantic mapping: a functional or relational definition that connects both, the elements of the syntax and the elements of the semantic domain.

This technique of giving meaning to a language is the basic principle of denotational semantics: every syntactic construct is mapped onto a semantic construct. As explained in [2] the semantic mapping has the form:

$$Sem : L \rightarrow \mathcal{P}(S)$$

and thus functionally relates any item in the syntactic domain to a set of constructs of the semantic domain. The semantics of a model $m \in L$ is therefore $Sem(m)$ denoting a set of elements in the domain $S$.

Given any two models $m, n \in L$ combined into a complex one $m \oplus n$ (for any composition operator $\oplus$ of the syntactic domain), the semantics of $m \oplus n$ is defined by $Sem(m \oplus n) = Sem(m) \cap Sem(n)$. This definition also works for sets of documents which allows an easy treatment of views on a system specified by multiple diagrams. The semantics of several views, e.g., several UML documents is given as $Sem(\{doc_1, \ldots, doc_n\}) = Sem(doc_1) \cap \ldots \cap Sem(doc_n)$. A set of models $docs$ is consistent if elements of $S$ exist that are described by the models, so $Sem(docs) \neq \emptyset$. As a consequence, the approach supports both view integration and model consistency verification.

In the same way, $n \in L$ is a (structural or behavioral) refinement of $m \in L$, exactly if $Sem(n) \subseteq Sem(m)$. Hence, refinement is nothing else than "$n$ is providing at least the information about the system that $m$ does". These general mechanisms provide a great advantage, as they simplify any reasoning about composition and refinement operators and also work for incomplete models.

**Semantic Domain.** We identify a single semantic domain $S$ used as a target for the semantic mapping of various kinds of modeling languages. Since we are interested in object-oriented modeling languages, the domain should provide concepts commonly found in object-oriented systems. The system model, first defined in [8] and extended in [9], defines these concepts. Generally, the system model characterizes object-oriented systems using basic mathematical theories. The semantics of a model $M$ is hence given as a set of all systems of the system model that are possible realizations of the model $M$. This way, we obtain an adequate and relatively easy to understand semantic domain which is crucial for the acceptance of a semantics definition.

To capture and integrate all the orthogonal aspects of a system modeled in, e.g., UML, the semantic domain necessarily has to have a certain complexity. Related approaches to UML semantics very often define a relatively small and specialized semantic domain and can therefore not capture the multitude of concepts typically found in a complex modeling language. More details on the system model are presented in Sect. 3.

**Tool Support.** Having the system model at hand, we could define the semantics of a language using pencil and paper. This was done for UML class diagrams [10] and Statecharts [11]. Tool support, however, is beneficial in two ways. First, we specify a machine-readable, checkable semantics that can directly be used for verification purposes. Second, the different artifacts can be better controlled and quality checked by using standard tools, e.g., version control.

Fig. 2 gives an overview of the default approach when defining the semantics of a language with tool support. First, the (domain specific) modeling language concepts are specified using a MontiCore grammar. MontiCore [6] is a framework for the textual definition of languages based on an extended context-free grammar format. This format enables a modular development of the syntax of a language by providing modularity concepts like language inheritance. Framework functionality helps developers also to define well-formedness rules and, for example, the implementation of generators.

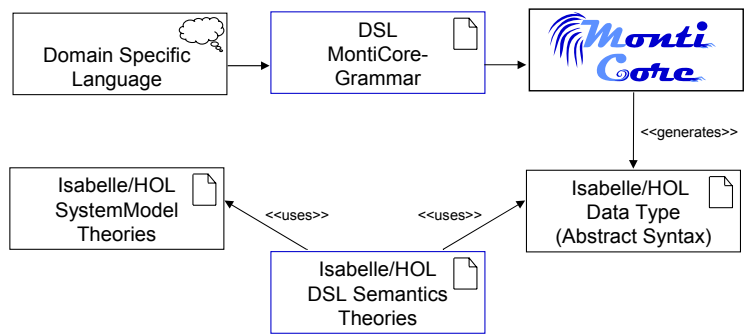

**Fig. 2.** Default Approach with Tool Support

To provide the semantics developer with maximum flexibility but also with some machine-checking (i.e., type checking) of the semantics and the potential for real verification applications, we use the theorem prover Isabelle/HOL for

- the formalization of the system model as a hierarchy of theories,
- the representation of the abstract syntax of the language as a deep embedding, and
- the semantic mapping which maps the generated abstract syntax to predicates over systems of the formalized system model.

The formalization of the system model as theories in Isabelle/HOL has to be done once and is described in Sect. 3. We have implemented a generator in MontiCore that produces an Isabelle/HOL data type representing the abstract syntax of the language, given a MontiCore grammar as input. Details on the derivation of the abstract syntax are explained in Sect. 4. The semantic mapping is also contained in Isabelle/HOL theories, an example is given in Sect. 5.

This approach of using a deep embedding has mainly two advantages over a shallow embedding. First, we can benefit from the sophisticated mathematical notation in Isabelle/HOL for defining the semantic mapping. Second, since both, the syntax and the mapping are formalized, we are able to reason about syntactic properties of concrete models and, more importantly, about properties of the mapping itself. This is in contrast to a shallow embedding where we would generate predicates directly from concrete models. This approach has some advantages when reasoning about concrete model properties but does not allow reasoning about the syntax or the mapping at all. Furthermore, we would have needed to invent another mathematical language to express the predicates outside Isabelle/HOL. As an extension to this approach (not shown in the figure), not only the abstract syntax data type is generated but also another generator that is able to translate concrete models to the abstract syntax representation as an instance of the generated data type. This is very useful when verifying properties of models and will be shown with the help of an example in Sect. 4.

**Handling Semantic Variations.** As mentioned in the introduction, the semantics of a modeling language should not be fixed but there should be explicit points where the interpretation of constructs can be specialized. These semantic variation points can be found in the system model but also in the semantic mapping or syntax. Variation points do not necessarily contradict interoperability: A comprehensive list of realization choices may serve tool builders as a definite reference when stating compliance to a given language.

In the system model [9] a large number of variation points has already been made explicit and different alternative configurations for variation points have been defined. Examples are the existence of multiple inheritance between classes, different realization strategies for associations, or different notions of type-safe overriding of methods. These semantic variations can be constrained prior to the semantic mapping but can also be left open.

For handling semantic variations in the syntax or in the mapping we propose to model these variations as stereotypes known from UML and to explicitly consider these stereotypes in the semantic mapping. The decisions of how particular syntactic elements should be interpreted can then be made by the modeler and need not be fixed beforehand. Additionally, there are dependencies between semantic variation points that have to be considered. A more complete account on how to handle semantic variations is however outside the scope of this paper.

## 3   System Model and Its Formalization

The system model is the universe of all possible object systems that can be modeled using an object-oriented modeling language like UML. It describes amongst other aspects the structural part of such systems, i.e., types, values, classes, objects and associations. Besides reasoning about the structure of systems it is also possible to specify or analyze behavior. The control part of the system model covers events and flow of information as well as execution of methods.

All systems are interpreted as timed or untimed global state machines (STS). Using the power of underspecification and variation points, the system model becomes very comprehensive but versatile in use. Due to space restrictions, we only present a small portion of structural definitions. To get a more complete picture of the system model features, the reader is referred to [9].

Main concepts of object-oriented modeling languages like types or classes appear in the system model grouped in corresponding universes, e.g., UTYPE or UCLASS. The universes contain only abstract identifiers. For example, classes are identified by elements of UCLASS and are only described by functions that yield information about their attributes, methods, or super-classes. They are never constructed from records or constructively represented in similar structures.

The system model itself is built in a modular and hierarchical way starting with a base theory about simple types and values. On top of this theory further theories define classes and objects as well as formalizations of the state of systems. The basic theories of the system model can be seen in Fig. 3.

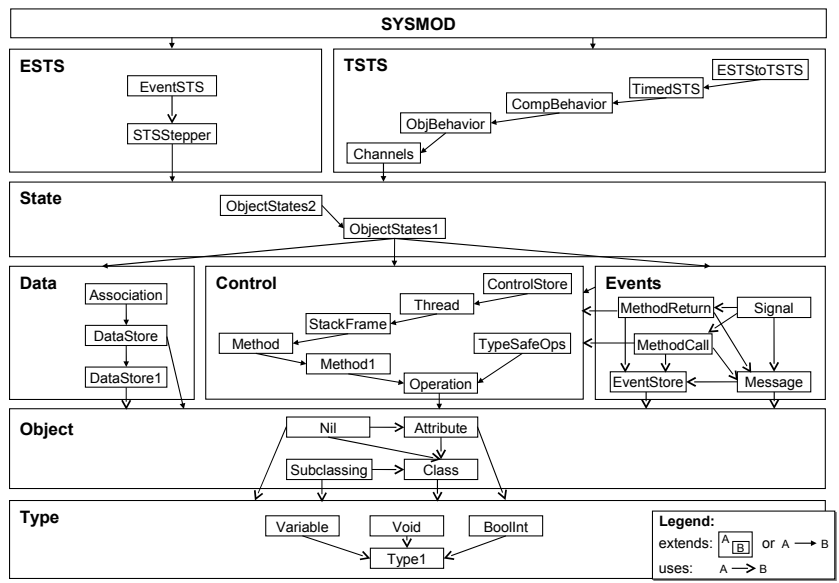

**Fig. 3.** Theories that constitute the system model

To support reasoning in the system model and the construction of the semantic mapping by tools, the proof assistant Isabelle is used for an implementation based on Isabelle/HOL. Isabelle's logic HOL [5] offers an implementation of functional programming and set theory.

Universes of the system model are implemented by corresponding data types since functions in Isabelle/HOL operate on data types. All introduced universes are universes of specific instances (systems) of the system model. They are retrieved from an instance of the system model with functions similar to selection functions on records [12]. For example, the function

```
consts UTYPE :: "SystemModel ⇒ iTYPE set"
```

maps systems to their universe `UTYPE` which is a set of type names from `iType`. Universes can comprise different sets of data type elements for concrete systems. An underlying data type (here `iTYPE`), is necessary since HOL sets have to be typed. Some universes contain others (e.g., $UCLASS \subseteq UTYPE$) which is modeled using wrapping constructors in the underlying type:

```
datatype iType = ...| TClass iClass | ...
```

The elements and sub-universes of a universe are defined as parametrized data type constructors yielding concrete values. This makes it possible to create and identify certain instances by names (here lists of characters). A class name in the system model can be created using the constructor `Class Name` of type `iClass` (with `Name = "char list"`). When reasoning about concrete instances this facilitates the creation and referencing of explicit names for elements.

All functions in Isabelle/HOL have to be total. Partial functions can be mimicked via the special type `a' option = Some a' | None` where `a'` is a type variable. Underspecified functions of the system model are introduced as constants of corresponding function types in Isabelle/HOL. Properties of functions or universes are given in definitions or predicates over systems. As an example, the underspecified function `CAR` is introduced as

```
consts CAR :: "SystemModel ⇒ (iTYPE ⇒ iVAL set)"
```

For every type name in `UTYPE` the function `CAR` yields all possible values an entity of the type can have in an instance of the system model. In its mathematical definition the function $CAR$ fulfills the property

$$\forall u \in UTYPE : CAR(u) \neq \emptyset$$

This is realized by the predicate `pCAR_Type1` that needs to hold for all valid systems `sm` of the system model:

```
pCAR_Type1 sm = (∀ u ∈ UTYPE sm . CAR sm u ≠ {})
```

The Isabelle implementation of the system model is split up in theories according to the structure in Fig. 3. Properties of system model instances as well as additional properties imposed by, e.g., variation points are declared in corresponding theories. These declarations have to be included as predicates when reasoning about instances. Predicates for elements of systems always have the signature `"SystemModel ⇒ bool"` (see Fig. 4 for a transitivity definition of the sub class relation) and thus are predicates on systems rather than on single functions or elements. This makes the combination and reuse of predicates much simpler.

```
1  constdefs pSubTrans :: "SystemModel ⇒ bool"
2  "pSubTrans sm == (
3    ∀ a b c . the (sub sm a b) ⟶ the (sub sm b c)
4                ⟶ the (sub sm a c))"
```

**Fig. 4.** Definition of a predicate about transitive sub class relations

To capture all systems `sm` ∈ `SystemModel` with non-empty universes *UTYPE* and a transitive sub class relation *sub* one has to write:

$$\{sm. \ \text{pCAR\_Type1} \ sm \ \wedge \ \text{pSubTrans} \ sm\}$$

The use and combination of theories and variation points is thus a partially manual composition task. This may be improved in the future when focusing more on the usability part of our implementation.

## 4 Concrete Syntax and Derivation of Abstract Syntax

In this section we briefly introduce MontiCore grammars to specify the syntax of a modeling language, explain its modularity concepts, and show how to derive the Isabelle/HOL abstract syntax data type. We present matters with the help of UML-like class diagrams as a shortened example sufficient to show the main concepts. Please note that the general idea can also be transferred to other tools that process context-free grammars or even metamodels.

In MontiCore, modeling languages are syntactically defined with context-free grammars like the one in Fig. 5. By language inheritance, the grammar re-uses productions of a super-grammar `Common` (l. 1) where, e.g., the commonly used non-terminals `IDENT` or `Type` are defined. The keyword `external` (l. 4) indicates that a second language for invariants is embedded. Later, this production can be mapped to any invariant language. Interface productions (l. 3) state that any implementing production (lines 10 and 18) can be parsed when the interface is expected (l. 8). Enumerations (l. 14) list possible alternative terminal symbols. Other than that, MontiCore grammars have terminal symbols enclosed in quotes (e.g., l. 7), alternatives (|), iteration (*), and optional elements (?). Fig. 6 contains two simple concrete models that conform to the grammar of Fig. 5.

The MontiCore generator basically derives a set of Java classes representing the abstract syntax and an ANTLR-based parser that can process the models from Fig. 6. For our purpose, an additional generator has been implemented that produces an Isabelle/HOL theory that holds the abstract syntax as a set of data type definitions, see Fig. 7. The theory imports data types generated from super-grammars and the theory that fills the language parameters for externals (l. 2). Recursively dependent types are computed and generated as a single data type (not shown in this example). Iteration is translated to the built-in type `list`, optional elements to type `option`. The interface `CDElement` leads to a

```
1 grammar CDSimp extends mc.umlp.common.Common {
2
3   interface CDElement;
4   external Invariant;
5
6   CDDefinition =
7     "classdiagram" Name:IDENT
8     "{" ( CDElement | Invariant ";")* "}";
9
10  CDClass implements CDElement =
11    "class" Name:IDENT ("extends" scl:IDENT ("," scl:IDENT)*)?
12    ("{" (CDAttribute)* "}" | ";");
13
14  enum CDModifier = "public" | "private";
15
16  CDAttribute = CDModifier? Type Name:IDENT ";";
17
18  CDAssociation implements CDElement =
19    "association" Left:IDENT "--" Right:IDENT ";";
20 }
```

**Fig. 5.** MontiCore grammar of class diagrams

```
1 classdiagram ABC {          classdiagram CA {
2   class A;                    class C;
3   class B extends A;          class A extends C;
4   class C extends B;        }
5 }
```

**Fig. 6.** Two simple class models

data type with alternative constructors, one for each implementing type (l. 18). Enumerations become types with an alternative constructor for each possible value (l. 5).

The generated theory now holds a deep embedding of the syntax in Isabelle/HOL and can be used to define the semantic mapping. Since we also want to be able to reason about concrete models, we also have to translate these to instances of the data type. For that purpose, our MontiCore generator additionally produces a specific generator that translates concrete models. Applied to the model **ABC** we obtain the theory shown in Fig. 8. The constant **abc** (l. 4) is a class diagram that has name **ABC**, an empty list of invariants, and three class diagram elements which are all classes. All classes have no attributes but some have a super-class, e.g., **CDClass ''C''** has super-class ''B''.

Please note that the generator actually produces separate constants for each data type. It has been in-lined here for the sake of brevity.

```
1   theory CDSimpAS
2   imports "$UMLP/abstractSyntax/external/ExternalCDSimpAS" CommonAS
3   begin
4
5   datatype CDModifier =
6       CDModifierPRIVATE
7     | CDModifierPUBLIC
8
9   datatype CDAttribute =
10      CDAttribute "CDModifier option" Type IDENT
11
12  datatype CDClass =
13      CDClass IDENT "IDENT list" "CDAttribute list"
14
15  datatype CDAssociation =
16      CDAssociation IDENT IDENT
17
18  datatype CDElement =
19      CDElementCDClass CDClass
20    | CDElementCDAssociation CDAssociation
21
22  datatype CDDefinition =
23      CDDefinition IDENT "Invariant list" "CDElement list"
24
25  end
```

**Fig. 7.** Abstract syntax data type in Isabelle/HOL

```
1   theory ABC
2   imports "$UMLP/abstractSyntax/gen/CDSimpAS"
3   begin
4   constdefs "abc == CDDefinition  ''ABC'' []
5     [CDElementCDClass (CDClass  ''C'' [''B''] []),
6     CDElementCDClass (CDClass  ''B'' [''A''] []),
7     CDElementCDClass (CDClass  ''A'' [] []) ]"
8   end
```

**Fig. 8.** Concrete model representation

## 5   Semantic Mapping and Its Formalization

All necessary components for semantic mappings are now available for the use in Isabelle: the language itself as a data type and a formalization of the system model. Functions in Isabelle/HOL can be used to define mapping functions from the implementation of the abstract syntax to the system model implementation. Features like recursion, constructor pattern matching and functional decomposition can be incorporated. The domain of the mapping function is the generated top-level data type of the language to be mapped. Its range is the power set of systems of the system model. Instances of systems are of the data type `SystemModel` in the Isabelle implementation.

From the UML class diagram grammar the data type `CDDefinition` in Fig. 7 is generated. The corresponding mapping function has the signature

$$\texttt{mCDDefinition :: "CDDefinition} \Rightarrow \texttt{SystemModel set"}$$

What the mapping function basically does is adding constraints to a set of systems. I.e., the mapping describes properties of the elements in its returned set of systems. Essential constraints are that every system has to fulfill a set of basic predicates like `pCAR_Type1` and `pSubTrans` from Sect. 3. This way the mapping only renders valid instances of the system model. Further constraints depend on the mapped modeling language. The mapping function can be decomposed to many short and compact functions each mapping one aspect of the abstract syntax. The function to map the data type `CDClass` (l. 15, Fig. 7) is shown in Fig. 9.

```
1  fun mCDClass :: "CDClass ⇒ SystemModel ⇒ bool"
2  where
3    "mCDClass (CDClass name supers attrs) sm = (
4    ∃ c ∈ UCLASS sm .
5        c = Class (mIDENT name) ∧
6        gall supers (mSuperClass c sm) ∧
7        gall attrs (mCDAttribute c sm)
8    )"
```

**Fig. 9.** Mapping of data type `CDClass`

This predicate on systems enforces that a class exists in `UCLASS` which has the specified class name and also fulfills further constraints given by the mapping of the super-classes `mSuperClass` and the mapping of the attributes `mCDAttribute`. The function `gall` feeds all elements of the list `supers` as a third parameter to the function `mSuperClass` which is called with the current class and system as parameters. These functional decompositions of the mapping make it easier to write comprehensible and maintainable code.

## 5.1   Example: Cyclic Inheritance Problem

To demonstrate the use of our implementation of the system model and the generation of instances from concrete models we present a short example. The textual models for this example were already given in Fig. 6. The semantic mapping renders a set of systems that fulfill the specifications given by the textual class diagrams. If a system complies to both specifications it is contained in the intersection of both mappings. Following the paradigm convention over configuration classes with same names in different systems share the same identity. Thus all systems in the intersection of the mappings contain a circular inheritance, i.e., `A extends C extends B extends A`.

```
1   constdefs pSubNonCirc :: "SystemModel ⇒ bool"
2   "pSubNonCirc sm ==
3       (∀ c1 c2 . (the (sub sm c2 c1) ∧  the (sub sm c1 c2)
4                    ⟶ c1 = c2))"
```

**Fig. 10.** Definition of a predicate for non-circular inheritance of classes

```
1   lemma SubNonCirc:
2       "⟦pSubNonCirc sm;the (sub sm c2 c1);the (sub sm c1 c2)⟧
3                    ⟹ c1 = c2"
4   by (unfold pSubNonCirc-def, auto)
```

**Fig. 11.** Rule to apply the predicate for non-circular inheritance

In this example we show a proof in our system model implementation that no system from the combined specification in Fig. 6 is compatible with the specification of non-circular inheritances given in Fig. 10. The lemma and the corresponding proof can be found in Fig. 12. The additional lemma in Fig. 11 is used to utilize the definition of `pSubNonCirc` in a more convenient way. The same is done for the definition `pSubTrans` from Fig. 4 in a corresponding lemma `SubTrans`.

```
1   lemma ABC-CA-circ: "mCDDefinition ABC.abc ∩ mCDDefinition CA.ca
2                        ∩ {sm . pSubNonCirc sm} = {}"
3   apply(unfold abc-def ca-def,auto)
4   apply(frule SubTrans, auto)
5   by(frule SubNonCirc,auto)
```

**Fig. 12.** Lemma and proof using generated UML models

First the definitions `ABC.abc_def` and `CA.ac_def` are unfolded replacing `ABC.abc` and `CA.ac` by their values (shown in Fig. 8 for the first model). To complete the proof the transitivity of the sub class relation (lemma `SubTrans`) is employed yielding that `Class ''A''` is a sub class of `Class ''B''`. Afterwards the rule `SubNonCirc` leads to `Class ''B'' = Class ''C''` which is an obvious contradiction here. Automatic simplification is done by the proof command `auto` throughout the proof.

In the example, two models of the same language are used. However, handling models of different languages is done in exactly the same way, since the semantics of a model is always given as predicates over the same type `SystemModel`.

## 6   Related Work

Quite a number of approaches to define a formal semantics for programming and modeling languages exist; a survey is given in [13,14]. These works deal with formalisms and mathematical frameworks that tend to be too complex or cumbersome to use for industrial applications. Efforts to bridge this gap led to reasoning tools to support using these formal/mathematical frameworks. Prominent works have shown that proof assistants can be used to define and verify semantics of programming languages [15,16,17,18]. But as discussed the execution semantics of programming languages is not directly suitable for underspecified modeling techniques.

Works around Java compilers [19] and virtual machines [20] show that the embedding of languages and the derivation of a proof environment are a tedious but crucial task. We automate the task of embedding modeling languages in a proof environment and offers the system model as a reasoning framework.

One of the earliest frameworks for designing and analyzing domain specific programming languages (DSPLs) is the CENTAUR system [21]. It is a combination of different tools to define the syntax, transformations and an expression evaluation and reasoning framework using Prolog and Coq [15].

Semantic anchoring is a more recent approach for defining semantics of modeling languages [22]. The semantics is defined based on semantic units which are minimal languages with well-defined semantics for models of computations. The abstract syntax of domain-specific modeling languages is transformed to the abstract syntax of a semantic unit. In [22] an example of semantic anchoring with tool support for defining and transforming models is given. The work also covers similar topics and tool support addressed in this paper but is primarily about giving operational semantics through generated AsmL [23] sources. Other approaches, e.g, [24] are based on MOF [25] for which formal semantics exist [26]. In [27] the authors propose a composition of semantic units when modeling heterogeneous systems that do not match a single semantic unit. The composition is not supported by tools yet. Heterogeneous UML semantics approaches such as [28] also use a posteriori composition of semantics. In our approach we circumvented this problem by starting with a powerful enough system model.

A completely integrated approach to define a formal language and its semantics is shown in [29]. The abstract syntax and static semantics of modeling languages can both be expressed in one Alloy [30] model. A major advantage is the integrated development of all parts of the language using only one formalism. Alloy relies on the small scope hypothesis and uses only a bounded search space to find counterexamples.

## 7    Conclusion and Future Work

The main contribution of this work is the provision of a flexible tool support for system model-based semantics definitions. The predicative semantic mapping helps us to cope with underspecified models. We provide the system model as a predefined and rather general semantic domain that can be reused in various semantics definitions for structural, behavioral and interaction concepts. Furthermore, the form of semantics definition based on sets allows for an easy explanation of composition and refinement of models.

The syntax and semantics can fully be defined using the tools MontiCore and Isabelle/HOL. Using a theorem prover allows us to define semantics in a very flexible and modular, yet machine-readable way. A MontiCore generator is used to deeply embed the abstract syntax of a language defined in MontiCore into Isabelle/HOL. Based thereon, also concrete models can be translated into Isabelle theories that provide means to directly use the semantics for verification purposes. The whole approach was shown for a simple example.

Using a theorem prover gives us great power and flexibility to handle all kinds of verification problems. But clearly, automation is rather poor compared to, e.g., model checking, since proofs have to be conducted manually. Future work will therefore be concerned with the question how to improve automation, e.g., by generating a set of helpful auxiliary lemmas and definitions. The identification, management, and consistent configuration of variation points has not been discussed in detail, this will be a matter of future work, too. Finally, we plan to further investigate which conclusions we can draw from the integrated semantics of languages, hoping to find new insights of how different languages interact with each other.